# Modeling Endogenous Social Networks: The Example of Emergence and Stability of Cooperation without Refusal

**David Chavalarias**[1]

Center for Research in Applied Epistemology (CREA)
Ecole Polytechnique / CNRS – www.polytechnique.edu,
1, rue Descartes, 75005 Paris, France
**David.Chavalarias@polytechnique.edu**
http://www.chavalarias.com
Tel : (+33) 1 55 55 83 26, Fax : (+33) 1 55 55 90 40

**Abstract**

Aggregated phenomena in social sciences and economics are highly dependent on the way individuals interact. To help understanding the interplay between socio-economic activities and underlying social networks, this paper studies a *sequential prisoner's dilemma* with binary choice. It proposes an analytical and computational insight about the role of endogenous networks in emergence and sustainability of cooperation and exhibits an alternative to the choice and refusal mechanism that is often proposed to explain cooperation. The study focuses on heterogeneous equilibriums and emergence of cooperation from an all-defector state that are the two stylized facts that this model successfully reconstructs.

[1] The author would like to thank T.K. Ahn, Paul Bourgine, and Jean-Louis Dessalles for helpful discussions and comments. This research was partly supported by the ACI "Systèmes Complexes en SHS - *Coalition, Marchés et Intractions*" and the Ecole Polytechnique, Paris.

*Introduction*

It is now fully admitted that aggregated phenomena in social sciences and economics are highly dependent on the way individuals interact (Cohendet et al. 2003, Föllmer, 1974, Schelling 1978 or Kirman 1983), i.e. on the social networks they form. Arguably, this dependence will strengthen, as technologies of information and communication will spread in all aspects of our daily life. Decentralized socio-economic interactions are now worldwide (groups on the Internet, e-mail, blogs, internet auctions, peer2peer, etc.) and at the same time traditional networks of relationships, trade, collaborations, etc., become physically implemented in softwares that enable users to manage large address books.

Besides physical networks that determine the possibility and conditions of interactions, activity networks (who is actually interacting with who) are emergent phenomena of socio-economic activity at the same time they influence this latter. People meet each other, have some kind of socio-economic exchange, learn from these interactions, adapt their future actions and fill address books that will influence their future partner selection. To help to understand the interplay between socio-economic activities and the underlying social networks, this paper takes the example of emergence of cooperation in a society of interacting agents and studies a *sequential prisoner's dilemma* in the framework of dynamics population games with binary choice. This work builds on the model of Ahn et al. 2003 and proposes an analytical and computational insight about the role of endogenous networks in emergence of cooperation. It proposes an alternative to the choice and refusal mechanism that is often proposed to explain cooperation.

## I. Endogenous networks in dynamic population games

A dynamic population game is a game played over time by large populations of boundedly rational players who interact in discrete time. The outcome of these interactions is defined by a payoff matrix *G* (symmetric most of the time). A good overview of dynamic population games can be found in Fagiolo 2004. In the following, we will consider only dyadic interactions with binary choice (cf. box 1).

---

**Box 1 : Dynamic population games with dyadic interactions and binary choice**

The standard framework for dynamic population games consists in a set of *N* individuals that play games in discrete time. At any time period t ≥ 1, each individual *i* plays a strategy $s_i \in S$. In binary choice games, $S=\{s_1, s_2\}$ and that whenever two agents (say *i* and *j*) interact, they play a bilateral game which (symmetric) payoff matrix (to agent *i*) is given by:

| Agent 2 → ---- ↓ Agent *i* | $s_1$ | $s_2$ |
|---|---|---|
| $s_1$ | a | b |
| $s_2$ | c | d |

with $(a,b,c,d) \in R^4$.

The set of agents a given agent *i* is able to interact with is called its neighborhood $V_i$ and at each time period, each agent is able to interact with some of its neighbors. Depending on the model under study, agents are able to adapt their strategy, their neighborhood or both, following some predetermined decision rules, often indexed to payoffs obtained in previous periods.

Dynamic population games are completely defined once one specifies:
(i) the payoffs in the matrix *G*;
(ii) the interaction structure in place at each given point in time (or the rule that governs how the interaction structure changes over time),
(iii) the set of decision rules and strategies.

It is very frequent that the outcomes of the game in the long run is not sensitive to a positive affine transformation of each player's payoffs' function (cf. Weibull, 1995). In that case, every 2×2 games with binary choice can be reduced to a game which associated matrix has only two free parameters.

---

In case of games on networks, two mains strands of modeling coexist. The first one considers games where the interaction structure is given and static. Modelers are then interested in the influence of the networks' structure on the population dynamics or on the existence and stability of equilibriums of the game. The second strand considers games where agents can endogenously adapt their local neighborhood by adding or deleting links with other agents (Cowan et al. 2003; Fagiolo 2005; Jackson & Watts 2002).

In the two strands, the most studied dynamic population games are *coordination games* and *the prisoner's dilemma game* (PD game). In this paper, we will focus on the latter to study a model of endogenous network formation. Thus, we will briefly survey the literature on prisoner's dilemma game on networks.

### 1.1 The prisoner's dilemma on networks

A prisoner dilemma game is a dynamic population game where the payoffs matrix has the following structure:

| Agent 2 →<br>----<br>↓ Agent 1 | C | D |
|---|---|---|
| C | r | 0 |
| D | 1 | p |

With the two conditions *0<p<r<1* and *r+p<1*.

The game can be played several times in a row (repeated PD) or only once (one shot PD), players can play simultaneously (original PD game) or sequentially.

The prisoner's dilemma game belongs to the class of social dilemmas and has become a paradigm for modeling cooperation, i.e. situations where one achieves actions that benefit to the other at its own expense. In its sequential form, the PD game or some of its variant can also be used as a scheme to study trust in social networks (Dasgupta 2000; Hayashi et al. 1999). In both situations, there is a dilemma between individual and collective rationalities. The dilemma lays in the fact that option *D* (defect) leads always to the highest reward whatever the other does. But when both players chose option *C* (cooperate), they receive *r>p*. This means that mutual cooperation is always more advantageous than mutual defection (collective rationality), but given the opponent's action, defection is individually more advantageous than cooperation (individual rationality).

Experimental studies on prisoner's dilemma, and more generally on social dilemmas typically find that, if all players are taken to care only about their own pecuniary reward, the level of cooperation observed is not consistent with non-cooperative game-theoretic equilibrium predictions (Clark and Sefton, 2001; Henrich et al., 2001; Schmidt et al., 2001, Watabe et al., 1996; Hayashi et al. 1999 for the sequential form). In particular, one of the most puzzling phenomena observed in experimental studies is that among the various observed behaviors, there is often a substantial proportion of altruist behaviors, i.e. unconditional cooperators (Ahn et al. 2003). The standard question addressed by modelers is then to explain how rational individuals can go beyond the dilemma and choose cooperation rather than defection or how they can trust each other. It is a fundamental question when it comes to understand the functioning of biological, social or economic systems and it has attracted interest from scholars since seminal works of Hamilton (1964) and Trivers (1971).

Researches in the field first focused on simultaneous PD and demonstrated that cooperation can emerge if players are allowed to play repeatedly for a sufficiently large number of periods in a row each time they are paired (Axelrod & Hamilton, 1981; Axelrod, 1981). However, when interaction sequences are too short, when agents may be rapidly changing their strategies, or when agents may be turning over, explaining cooperation is slightly subtler (Hammerstein 2002). Moreover, it does not explain altruism. For these reasons, various mechanisms have been proposed among which indirect reciprocity (Nowak and Sigmund, 1998, Milinski et al. 2001), group selection (Wilson and Dugatkin, 1997; Pepper and Smuts 2000), indirect evolution (Güth & Yaari1, 1992), theory of prestige (Dessales, 1999), costly signalling (Smith et al. 2000), strong reciprocity (Bowles and Gintis, 2004), tags (Riolo et al. 2001) and altruistic punishment (Henrich and Boyd, 2001; Fehr & Gätcher, 2002).

All these approaches have in common that players are randomly matched. There is no physical network underling interaction and since there is no partner selection, there is no emergent activity network either. However, as researches in sociology and economics highlighted the importance of social networks (Granovetter 1973, 2005; Jackson 2003; Milgram, 1967; Travers & Milgram 1969), scholars begun progressively to study games on networks and new explanations of cooperation were proposed which advantage is often to require less sophistication on the part of individual agents (Nowak and Sigmund 2000).

Nowak and May (1992) showed that a population payoff-biased mimetic agents with binary choice and no memory placed at the nodes of a toric grid can sustain a positive level of cooperation in a certain domain of the parameter space, yet very limited (Hauert 2001). The originality of their paper lays in the fact that their model exhibits attractors with mixed population of *C* and *D* players. These results where extended to other forms of strategy updating in subsequent papers (Nowak and May 1993) and several papers explored the influence of network's topology, among which small worlds (Masuda & Aihara 2003), random graphs (Durán and Mulet 2003; Watts, 1999) and disordered networks (Tomochi, 2004)

Many variations have been proposed on prisoner's dilemma on networks exploring the influence of the action updating (Huberman et Glance, 1993), the rule for strategy selection (Szilagyi 2003) or the strategies' space itself (Cohen et al. 2001; Ebel & Bornholdt, 2002; Lindgren and Johansson, 2002; Lindgren 1995 to cite a few). If most models can explain sustainability of cooperation, few of them can explain both emergence of cooperation from an all defecting state and heterogeneous population as outcomes of the dynamics. However, one of the main conclusions of these studies is that the topology of the networks matters and more precisely, that cooperation is favored by a high degree of cliquishness of the networks of activity. This is a motivation to consider models with partner selection where the network's topology is an emergent property. What are the conditions under which local partner selection generates networks that can sustain cooperation?

Smucker et al. (1994) proposed a PD game with choice and refusal where the network is not given a priori. Players, defined by a finite state machine, have expected payoffs concerning interactions with some other players. They select their partners in function of this expected payoffs and can refuse to interact with a player if the expected payoffs associated are below a given threshold. Thus a network of who is playing with who emerges which affects individual payoffs. These payoffs are used to update individual strategies (the parameters of the machine) following and evolutionary dynamics. The authors show for small populations that cooperation is enhanced by this mechanism compared to the round-robin matching situation. Several models of partner selection with choice and refusal have been proposed since (De Vos and Zeggelink, 1997; Batali and Kitcher; 1995; Ashlock et al. 1996; Hruschka and Henrich, 2005; Zimmermann and Eguiluz 2003). Again, the main conclusions of these studies is that selection of interaction with choice and refusal enable cooperators to interact preferentially with other cooperators thus giving the network the appropriate level of cliquishness to support cooperation in some domain of the parameter space. However, we can notice a drift between the conclusions of these studies and the initial goals of researches on cooperation: here we explain mainly how conditional cooperation (through choice and refusal) can emerge whereas first studies on cooperation aimed to understand altruism i.e. unconditional cooperation.

### 1.2 Endogenous networks and cooperation without refusal

Ahn et al. (2003), proposed a prisoner's dilemma game with endogenous network formation through partner selection without refusal.

The main stylized facts this model aims to reconstruct are the endogenous emergence of a social network and the emergence and sustainability of cooperation in a heterogeneous population. Behaviors observed along several experimental studies can be roughly classified in three main types of strategies: selfish players, who most of the time defect, conditional cooperators, who cooperate if the other player did, and altruists players, which most of the time cooperate whatever the other did. The first aim of the authors was to explain how altruism (unconditional cooperation) could be a sustainable behavior in social network, even in presence of selfish agents.

The model is a sequential PD game in the framework of the indirect evolutionary approach (Güth and Yaari, 1992; Güth and Kliemt, 1998). It differs from existing models on several aspects. First, it doesn't introduce refusal mechanism, which is the main factor to explain cooperation. Second, as in Hruschka and Henrich (2005), it gives a special importance to errors in the choice of a partner, a factor that as a negative influence on the level of cooperation. It neither assumes that players can communicate about reputation of other players, which is a frequent assumption to explain trust in networks (Buskens 1998). Indeed, considering reputation brings about the problem of cooperation at the meta-level of information transmission. Last, this model tried to be as parsimonious as possible concerning the memory capacity of the agents: players cannot remember every interaction, which would be cognitively implausible, but only some of the most relevant ones.

In this paper, we give an analytical approximation of the dynamics of the model of partner selection proposed in Ahn et al. (2003). Then we validate this analytical insight by a comparison with computational studies. This enables to give the conditions for emergence of cooperation and identify the area in the phase space that supports heterogeneous populations. We also use this example to clearly highlight the role played by the endogenous networks in the emergence of phenomena such as cooperation and we propose an alternative interpretation for cooperation in networks in terms of higher social activity for cooperators rather than in terms of high cliquishness within groups of cooperators.

## 2. The model: how to use an address book

### 2.1 Definition of the game and strategies

Consider in discrete time a population of agents playing a *sequential PD* game as defined in *1.1*. Because the game is sequential, second movers know the action of their partner and can react accordingly.

We will assume that at each period, each agent has the opportunity to play only once as a first mover. With a probability of *(1-e)*, $e \in [0,1]$, first movers can choose their partner taking advantage of what they have learned from precedent interactions. With a probability *e*, first movers are matched at random with an unknown agent in the population. *e* can be interpreted as an error level in the partner selection process. We consider a complete graph for the network of accessibility (first movers can choose any agent in population), and we have an emergent directed graph for the network of activity: who is actually interacting with whom as a first mover.

Agents keep in an address book the names of partners of all their preceding successful interactions as first mover *((C,C) or (D,C))*. For a given agent *i*, names in *i*'s address book are called its *relations* and the set of agents that have *i*'s name in their address book are called its *friends*.

The game during one period proceeds in two steps (fig.2):

1) Each agent plays does a proposition to a partner it has eventually chosen in its address book (with probability *1-e*), playing *C* (cooperation) or *D* (defection).

2) Each agent answers as a second mover to *every agent* that has chosen it, *after* being informed of its moves (*D* or *C*). Consequently, an agent can play several times as a second mover but only once as first mover. We will check in the following that the number of second mover interactions remains plausible and compatible with time and cognitive constraints (cf. figure 4).

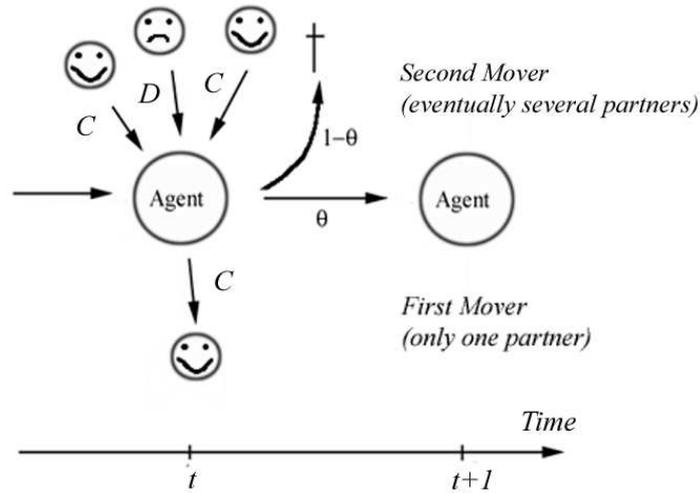

*One period time activity for one agent*

**Figure 1. Activity of one agent during one period.** Agent *i* first chooses a partner to play with as first mover, and proposes a deal (*C* or *D*). In a second step, *i* answers individually as second mover to every agent that chose *i* as partner. Between two periods, there is a probability *(1- θ)* the agent dies or leaves the game. In that case, *i* is replaced according to a replicator dynamics by a new agent with an empty address book.

The set *S* of strategies is composed of the three types mentioned previously: altruist (in proportion $\gamma$), reciprocator (in proportion $\delta$) and selfish. Although they are stylized types, we hope this set can give some insight on the phenomena at stake in social dynamics. These strategies are defined by their actions as first mover and as second mover as follow (Table 1):

**First mover:**

We will assume that agents know the proportions of the different types in population so that they can compute their expected payoffs[2]:

$E(C)=(\gamma+\delta).r$ ; $E(D)=(1-\delta-\gamma)p+(\delta+\gamma)$

The condition for having *E(C)>E(D)* is $\delta>p/r$ and first mover's strategies are defined as follow :

*a) When choosing a partner from address book (if not empty):*

---

[2] In a more sophisticated model, these proportions can be learned through interactions.

1. Altruists and reciprocators choose a name at random in their address book and cooperate.

2. Selfish agents choose an altruist at random in their address book and defect. If they don't know any altruist but have relations with an ambiguous status, i.e. altruist/reciprocator, they choose one at random and defect. This situation happens only when selfish agents cooperate as first mover, i.e. when $\delta > p/r$. When they know only reciprocators, they choose one at random and cooperate.

*b) When matched with an unknown partner:*

1. Altruists always cooperate as first mover (they are defined as unconditional cooperators).

2. Selfish agents and reciprocators maximize their expected payoffs: they cooperate if and only if $\delta > p/r$.

**Second mover:**

1. Altruist agents are unconditional cooperators. They cooperate (*C*) even if the first mover defected.

2. Reciprocators are conditional cooperators. They cooperate as second mover only if an only if the first mover did.

3. Selfish agents always defect (*D*) as second mover.

| Interaction | First mover | | Second mover (with friends) |
|---|---|---|---|
| | In address book (with relations) | Random | |
| **Selfish** | Search for altruist and defect. If none cooperate | Cooperate if $\delta > p/r$. | Defect |
| **Reciprocator** | Pick at random and cooperate | Cooperate if $\delta > p/r$. | Do what the other did. |
| **Altruist** | Pick at random and cooperate | Cooperate | Cooperate |

*Table 1* : behaviors of agents depend on the composition ($\gamma, \delta$) of the population.

Each time an agent has a successful interaction with an unknown partner, it adds the partner's name in its address book. The network of activity designed by the set of all address books is thus an evolving directed network. Agents begin their social life without knowing anybody, and little by little, they do friends and relations.

At the end of a period, the payoffs of agent $i$ are the sum of the payoffs of all $i's$ interactions at this period. These payoffs are assumed to be linked to a reproductive success: agents have a probability *(1-$\theta$)* to die or to leave the game after each round. Each died agent is replaced by a new agent according to a replicator dynamics so that the population size is kept at a constant size *N*: one agent is chosen in the population with a probability proportional to its payoffs at the preceding round, and a new agent with an *empty address book* of the same type is created. We have here an overlapping generation game.

The replicator dynamics is a standard in the modeling of social networks dynamics. We will not discuss its relevance to model social phenomenal like cooperation (see Chavalarias 2005 for example). The aim here is rather to see whether it is possible to tackle social dilemmas in situations where cooperation is a dominated strategy. Our hypothesis is that the replicators dynamics is a suited framework to identify some cultural phenomena that could counterbalance the advantage of defection on cooperation.

## 2. Analytical and computational studies of the model

### 2.1 The fitness function

We can see from table 1 that behaviors of agents varies according to the sign of $\delta$-$p/r$. The area of the simplex concerned by the problem of emergence of cooperation has to contain homogeneous populations of selfish agents (the point *(0,0)* in $\Delta$ ). It is then defined by $\delta < p/r$. For sake of clarity, we will give in this paper only the mean payoffs $\pi_i$ in this area. Complements can be found in the appendix.

**Proposition 1:** With the approximation that the proportions of types in a population can be considered constant at the agents times scales, the mean expected payoffs $\pi_e$, $\pi_r$ and $\pi_a$ for selfish agents, reciprocators and altruists in a population characterized by the proportions $\gamma$ of altruists and $\delta$ of reciprocators in the domain $\delta < p/r$ are given by the following equations (System 1):

$$\pi_{selfish} = (1-e)\left(\frac{2p(1-\gamma+\theta(\gamma-1))+\gamma}{1+\theta(\gamma-1)} + \frac{(1-\theta)\gamma}{1-(1-(\delta+\gamma))\theta}\right) + 2e(p(1-\gamma)+\gamma) \quad \text{(Eq. 1, } \pi_s\text{)}$$

$$\pi_{reciprocators} = (1-e)\left(\frac{\theta\gamma(r+2p-1)+\gamma+2p(1-\gamma-\theta)}{1+\theta(\gamma-1)} + \frac{r\gamma}{1-(1-(\delta+\gamma))\theta}\right) + e(2p+\gamma(1-2p+r)) \quad \text{(Eq. 2, } \pi_r\text{)}$$

$$\pi_{altruists} = (1-e)\left(\frac{r\theta\delta}{1+\theta(\gamma-1)} + \frac{r(2\gamma+\delta)}{1-(1-(\delta+\gamma))\theta}\right) + er(\delta+2\gamma) \quad \text{(Eq. 3, } \pi_a\text{)}$$

Proof is given in appendix. •

From this payoffs expression, it is easy to compute the relative fitness of each type given by:

$$f_i = \frac{\pi_i - \sum_j \pi_j}{\sum_j \pi_j} \quad \text{(Eq. 4)}$$

With this fitness function, it is now possible to study the dynamics of the population in the simplex $\Delta = \{0 \leq \delta, 0 \leq \gamma, \delta+\gamma \leq 1\}$ and find conditions for emergence and sustainability of cooperation. Whether cooperation will develop in the population will depend on the values of $e$ and $r$ and $p$. For sake of convenience and clarity we will adopt a standard parameterization of the PD game matrix taking $r=(1-p)$. This enables to consider a model with only two free parameters $e$, and $p$, which is convenient for visualization. Intuitively, $p$ fixes the strength of the social dilemma: the higher $p$ the stronger the dilemma.

Finding analytically the attractors of the dynamics defined by this fitness function is a mathematically hard task and for now, we will only find these attractors and predicted trajectories solving these equations numerically. However, there is already something to say analytically about potential paths toward cooperation.

*2.2 The unique scenario for emergence of cooperation*

From the equations above, we can begin to see whether it is possible that cooperation emerges in a population of selfish agents after introduction of few altruists. To do this, we have to look at the relative fitness of altruists in a population of selfish agents i.e. the sign of $\pi_{altruists} - \pi_{selfish}$ at the point $(\gamma, \delta) = (0,0)$.

**Proposition 2:** If selfish and altruist are the only available strategies, an all-selfish population is an evolutionary stable state (*ESS*).

**Proof :**

If we develop $\pi_{altruists} - \pi_{selfish}|_{\delta=0}$ in power of $\gamma$ at the origin we find:

$$\pi_{altruists} - \pi_{selfish} = -(1-e)\frac{2p(1-\theta)}{1-\theta} - 2ep + (1-e)\frac{\theta}{1-\theta}\gamma + o(\gamma) \quad (Eq.\ 5)$$

The first coefficient of this development is always negative and the second always positive. This means that a population composed by only selfish agents is an evolutionary stable attractor in the selfish/altruist space. •

The consequence of proposition 2 is that emergence of cooperation is never possible through introduction of few altruists in a population of selfish agents (except in the trivial case *p=0*). Nevertheless it often can maintain for some values of *e* and *p* once this proportion passes a certain threshold. On the other hand, in a population composed of a majority of selfish agents and few reciprocators ($\delta < p/(1-p)$), there is a neutral evolution of reciprocators among selfish agents with no cooperation ($\pi_r - \pi_s|_{(\gamma=0)\ and\ (\delta<p/(1-p))}$). The study of the fitness expression in the domain $\delta > p/(1-p)$ shows that above this threshold, reciprocators are sustainable even if they are first mover cooperators. Since this threshold is quite high, and given that neutral evolution is a very slow process, cooperation has little chance to emerge this way.

Consequently the only plausible scenario for emergence of cooperation starting from an all-selfish population in this model is to begin with a slow neutral evolution from a mixed population of selfish and few reciprocators. Then, when there are enough reciprocators, altruists could eventually be sustainable and grow quickly. This is a two steps evolution toward cooperation and these three types are consequently a *minimal set* for emergence of cooperation from an all-selfish population in presence of altruists.

Before we study the attractors defined by proposition *1,* we will confront analytical predictions to simulations.

### 2.3 *Convergence between analytical and computational results*

We present here a computational study that we compare with the predicted analytical trajectory (fig. 2). We have implemented in a multi-agents system the model presented above. We then ran series of simulations for a particular parameter setting. For each simulated evolution, we collected data on type's proportions in the population, address books and payoffs. We took here $\theta=0.99$, $e=0.3$ and $p=0.3$. Initial conditions are *10%* of altruists and *10%* of reciprocators for a population of *500* agents. As expected from the analytical study (circles on the graph), the population converges toward a mixed population of altruists and reciprocators and a full cooperative state. Cooperation has become sustainable and selfish agents have disappeared.

We can also verify that the stationary environment approximation made in proposition *1* is reasonable relatively to this example: the time between two markers (circles) represents one generation. The proximity of two successive markers indicates that the environment of an agent does not change much during its lifespan.

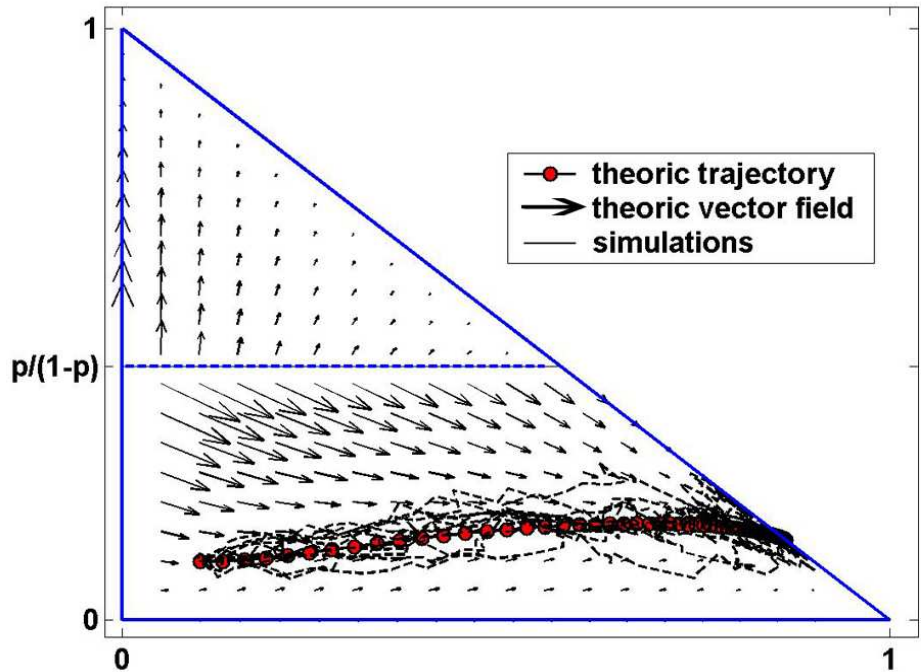

**Figure 2. Comparison between analytical predictions and simulations**. For $e=0.3$ and $p=0.3$, $\theta=0.99$ (simulation n°1) the theoretical analysis predicts very well the behavior of the system. The line with circles represents the theoretical trajectory in case of infinite population size. Dotted lines represent data from *10* independent simulations with *1000* agents each. In both cases, the community starting with a low rate of cooperation (80% of selfish agents, 10% of altruists and 10% of reciprocators) evolves toward cooperation. Differences between simulations and theoretical trajectory are mostly due to stochastic variations since for finite population the variance is not *0* as it can be seen on the mean payoffs plot of *fig. 4-b*. Each circle is separated by *100* periods i.e. the agents' mean lifetime expectancy.

The difference between theoretical results and simulations is mainly due to the "infinite population" approximation of the analytical model as it a can be seen on figure 3. When the system is close to the edge of the simplex (when at least the size of one population of one of the three types is very small), the mean empirical payoffs' distribution of the different types presents large variance and these means can be quite different from theoretical predictions in case of infinite populations. We can notice passing that the mean payoffs of each type increases as cooperation develops in the society.

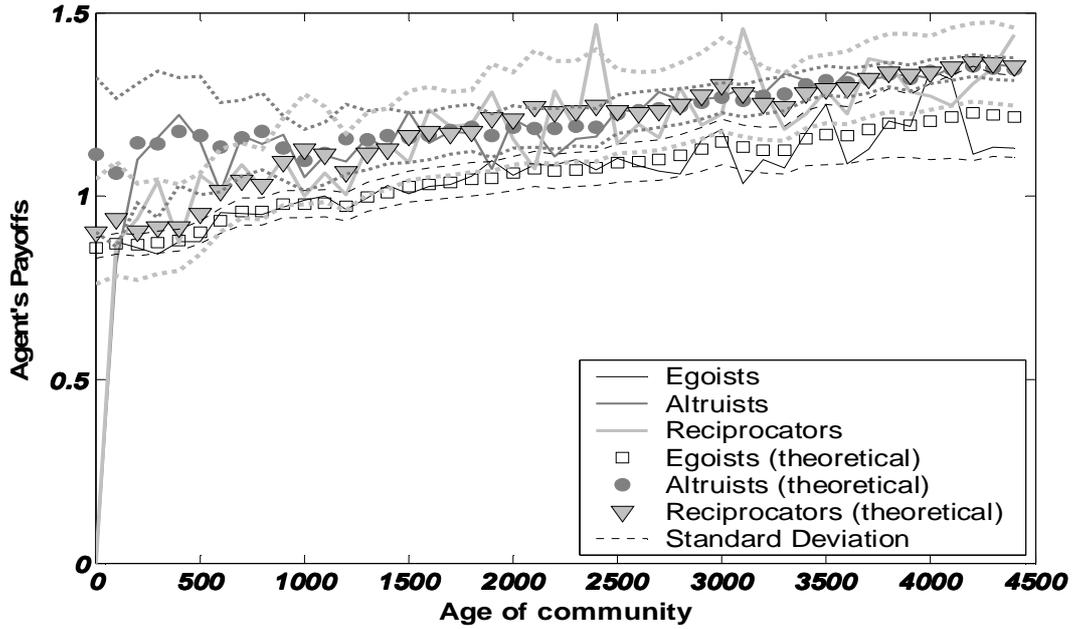

**Figure 3. Comparison between theoretical payoffs and empirical payoffs.** Dotted lines represent the theoretical standard deviation. We can see that the standard deviation for each type evolves respectively with the size of each sub-population. High variance of payoffs in small populations is the main source of differences between theoretical and empirical trajectory in *fig. 2*.

From our analytical study, we can derive the theoretical activity for each type in function of $\delta$ and $\gamma$ *(*see appendix*)* in order to see which type of agent is the most involved in social life. For this purpose, we defined the relative activity for altruists and reciprocators, given *(e,p)*, as the ratio between activities of these types and activity of selfish agents in function of *($\gamma,\delta$)* (equations are given in appendix), *i.e.*

a) The mean number of interactions per round for altruists on the mean number of interactions per round for selfish agents and

b) The mean number of interactions per round for reciprocators on the mean number of interactions per round for selfish agents.

We can observe on figure 4 that:

- These two ratios are always greater than one, which means that selfish agents are the less active in social network (they don't have friends actually).
- Altruists are the most active of all types, especially when in small proportion.

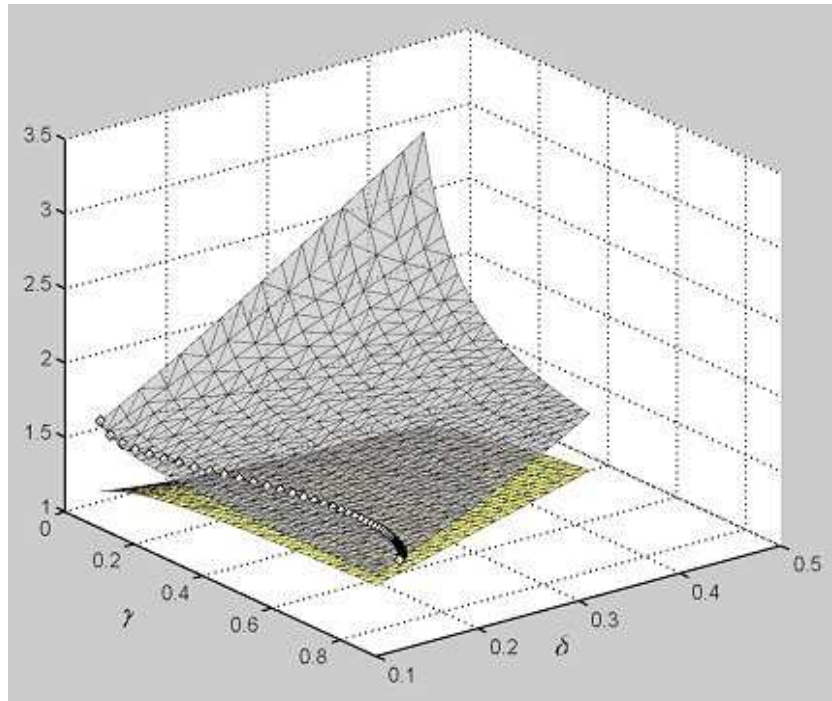

Figure 4: Theoretical relative activities for reciprocators and altruists ($0.1<\gamma<0.9$ and $0.1<\delta<0.4$). We can see that altruists are the most active of all types, especially when in small proportion. Here $e=0.3$ and $p=0.3$, $\theta=0.99$ (similar to simulation *1*) and the line plot is the projection of the theoretical trajectory presented in fig. 2 on the surface representing the relative activity of altruists agents. We can see that the difference between the activities of the different types is quite reasonable.

This example illustrates the way altruists can survive in presence of selfish agents: they are more active than everybody else. Since altruists cooperate first move, they can also do relations with reciprocators and consequently they do relations quicker. Altruists are more popular. The fact that their selfish partners cheat them is balanced by the fact that they are more appreciated and consequently more demanded as second mover partners in the social network

### *3. Looking for heterogeneous equilibriums*

Now that we checked that analytical predictions fit relatively well the dynamics of our model, we can look if this model is able the exhibit the stylized fact we presented in the introduction: stable heterogeneous populations of cooperators and defectors.

Given $(e,p,\theta) \in\ ]0,1[ \times ]0, 0.5[ \times ]0,1[$, the replicators dynamics is determined by the vector field $V$ in the simplex $\Delta$ which is given by its two components:

$$V_\delta = \frac{\delta \pi_r}{(\gamma \pi_a - \delta \pi_r - (1-\gamma-\delta)\pi_e)} - \delta \Big|_{e,p}$$

$$V_\gamma = \frac{\gamma \pi_r}{(\gamma \pi_a - \delta \pi_r - (1-\gamma-\delta)\pi_e)} - \gamma \Big|_{e,p}$$

A point $(\delta,\gamma) \in \Delta° \cap (]0,1[ \times ]0, \frac{p}{1-p}[)$ is an internal attractors if and only if

$$V_\delta^2 + V_\gamma^2 = 0 \Big|_{e,p} \qquad \text{(Equ. 5)}$$

We solved numerically *Equ. 5* to obtain the set of internal attractors for a given value of $\theta$. Figure *5-a* shows the set of couples *(e,p)* for which the dynamics presents an internal attractor. We can observe that this set has a non-empty interior. For $\theta=0.99$ it is crab's claw shaped and oriented along the line $p=0.5-0.35e$. As shown on figure *5-a*, the rate of cooperation at the attractor decreases in the population along this line, which means that $e$ has a negative impact on cooperation. Figure *5-b* shows the set of couples $(\gamma,\delta)$ that are internal attractors for some value of parameter *(e,p)*. On the graph, this set is discrete because the numerical algorithm that solved Equ. *1*. used discrete points. The actual set is the convex hull of each component appearing in figure *5-b*. We can see that most of these attractors are situated above the line *y=x* which means that there are generally more reciprocators than altruists at an internal attractors. This confirms our intuition from experimental studies.

Again, we can check that these analytical predictions fit the actual dynamics of the model. For example, let's look at the dynamics for the particular case $\theta = 0.99$, $e=0.7$ and

*p=0.25* (the arrow on fig. 5-a). Here the error level is quite high since agents have *70%* of chance to interact with an unknown partner. Equation *1* predicts an internal attractor at *(γ,δ)=(0.26, 0.23)*. With this parameter, a population starting after a neutral evolution, with *80%* of selfish agents, *19%* of reciprocators and *1%* of altruists actually converges toward an internal attractor, and cycles around it (*fig. 6*). With a finite population size (here *1000* agents), the population cycles around the attractor for the all the duration of the simulation (25 000 periods, *250* generations). As we can see, the presence of an internal attractor is well predicted by *Equ. 1* and solution of *Equ. 1* is close to the actual attractor. Asymmetries in the vector field's strength around the theoretical attractor as well as variance effects might explain the discrepancy between simulation results and predictions.

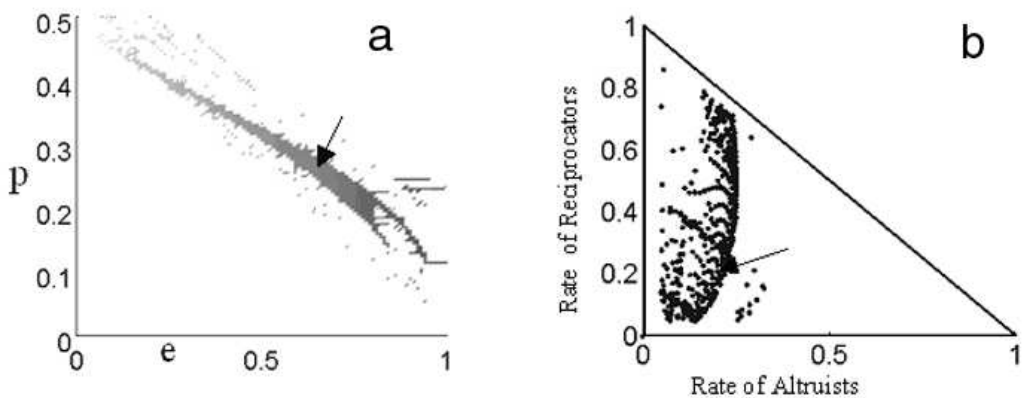

**Fig 5-a: Existence of internal attractors in the *(e,p)* space**. The color is indexed on the rate of cooperation at the attractor (the darker, the less cooperative). This has been obtained from *Equ. 5* solved by numerical analysis for *θ=0.99*. The arrow points to the couple *(e,p)* corresponding to the simulation presented in *fig. 6*. **Fig 5-b : Localization of attractors in the (γ,δ) space** obtained from numerical solution of *Syst 1*. Each point represents an attractor for some value of *(e,p)*. We can notice that for most attractors, there are fewer altruists than reciprocators, as observed in experimental settings. The arrow points to the couple (γ,δ) corresponding to the attractor of simulation 2 (*fig 6*).

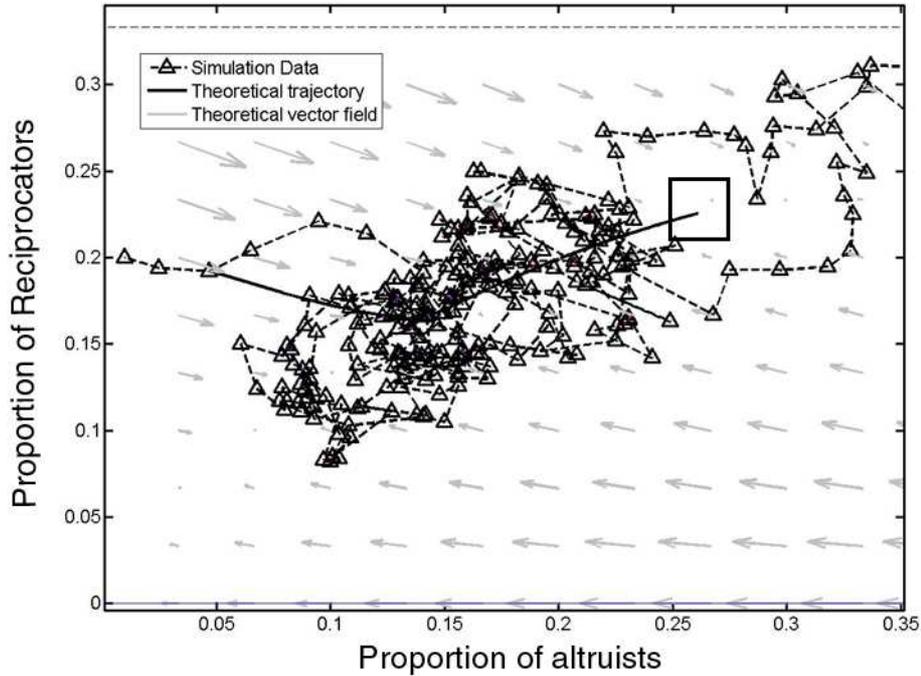

**Figure 6. A case of internal attractor** (*e=.7, p=.2, θ=.99*, simulation n°2) : For some values of *e* and *p*, a society starting after a neutral evolution from a defecting state (*19% of D-first-mover reciprocators, 80% of selfish agents, 1% of altruists* ) evolves toward an internal attractor with a mixed population of the three types. The simulation presented here is *25 000* periods long, corresponding to 250 mean lifetime expectancy. The black line represents the theoretical evolution in an infinite population and the theoretical attractor is situated in the black square. With finite population size (here *1000* agents), the population cycles around the attractor. Asymmetries in the vector field's strength around the theoretical attractor as well as variance effects might explain the discrepancy between simulation results and predictions.

For parameters that enable internal attractors it is particularly interesting to look at the structure of the emerging social networks to understand the interplay between network of activity and emergence of cooperation. Is the classical explanation in terms of cliquishness still relevant to explain cooperation in this model? Do altruists have most of their interactions with altruists? Who are the friends of most popular cooperators? It's enough to look at the components of the activities equations to see that the cliquishness explanation has to be discarded. Actually, cooperation can be sustainable even if altruists have more interactions with selfish agents and with reciprocators than they have with altruists.

This has an important impact on the network structure. In directed networks, important vertices for the structure description are often defined in terms of hubs and authorities (Kleinberg 1999). If a vertex points to many vertices with large authority weight, then it should receive a large hub weight. If many vertices with large hub weight point to a vertex, this latter should receive a large authority weight. In our model authorities are agents with a lot of friends and hubs are agents with good relations. If altruists were successful because of their cliquish behavior, we would expect that some of them should be both hubs and authorities. If we look at the details of the activity equations, we can see that while altruists are more likely to be authorities, selfish agents and reciprocators are more likely to be hubs. This emphasizes the importance of having the three types of agents to get behaviorally heterogeneous equilibriums: altruists can survive because they have a lot of friends and among them, reciprocators and altruists that guaranty a subsequent proportion of cooperative interactions; reciprocators can survive despite they have less friends than altruists because they avoid being exploited by selfish agents, selfish agents can survive because they exploit altruists.

We can see it very clearly if we plot the network of activity for a particular simulation with internal attractor: two agents are linked if one is the friend of on other. Figure 7 represents the ten hubs and ten authorities of a *100* agents population for *θ=0.99, e=0.7* and *p=0.25*. The ten authorities are all of altruist type while the ten hubs are all of selfish or reciprocator type. This kind of organization of the social network is very different from the one to be expected with models explaining cooperation with cliquishness. Because it is very characteristic, it could easily be identified in empirical data and future work should be able to determine whether a given cooperative social situations is more suitably described by this kind of model, by models with choice and refusal or by an other class of models.

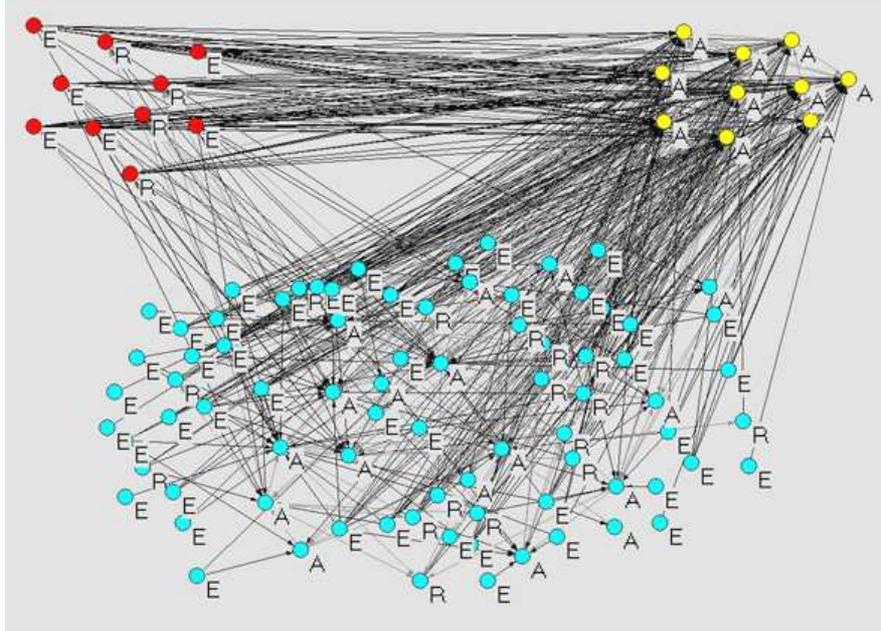

Figure 7. An example of network structure in the case of an internal attractor (*100* agents, *θ=0.99, e=0.7* and *p=0.25*). Nodes are agents. Arrows indicate who is the friend of whom. The graph shows the ten hubs and ten authorities of the network (a Pajek plot). Authorities are all of altruist type (*A*) while hubs are all of selfish (*E*) or reciprocator (*R*) type. This kind of organization (no overlap between hubs and authorities) can easily be identified in empirical or experimental data and is very different from the one expected with models explaining cooperation with cliquishness.

## 4. Conclusions

### *4.1 Network/population co-evolution*

We used the well-known paradox of the prisoner to illustrate the importance of the endogenous character of social networks for the understanding of social phenomena. While emergence of cooperation in biological or social systems has puzzled the scientific community for many years, introducing endogenous network enable to explain cooperation in heterogeneous social networks without the need of global representations of the society's structure nor high-developed cognitive capacities. All what are needed are local information and a limited memory on past interactions.

In such models, agents modify the network according to their types and in the same time the network configuration determines the fitness of the different types of agents in the population. This co-evolution between agents' types and networks' patterns leads to characteristic networks' structures that can eventually support a heterogeneous equilibrium.

We expect that this kind of co-evolution is crucial for the understanding of patterns emergence in social networks.

## *4.2 Emergence of cooperation, a two steps evolution?*

This study was also an opportunity to propose a scenario for emergence of cooperation, and not only for sustainability of existing cooperation. This scenario explains emergence of cooperation as a two steps evolution, the bottleneck being a neutral evolution phase between selfish agents and reciprocators.

This model is not human specific and could also be suited to explain cooperation in animal societies. Nevertheless, humans are much more efficient in classifying their partners in different types, keeping in memory past interactions and selecting intentionally their partners, leading to higher rate of cooperation.

We studied the role played by the strength $p$ of the social dilemma and the level $e$ of errors or instability of the environment. According to the particular nature of what is at stake in the social exchange, this model can be developed and adapted to describe different social phenomena. But we can already do some quick comparisons with sociological studies. For example, in small size societies, almost all interpersonal relations are direct ones. The population size is small; people know well each other and interact most of the time with known people. Relations are stable and cooperation can flourish. This situation corresponds to a low $e$ and is illustrated by simulation 1.

On the other hand, in situations where the environment is instable, agents have to interact often with unknown partners, and mechanisms of emergence of cooperation are more complex. This model explains this phenomenon as a two steps evolution toward cooperation starting from an all-selfish population (simulation 2). The key point is the possibility of a "neutral evolution" (selfish-reciprocators) followed by a directed one (spread of altruist strategies).

### 4.3 Popularity not cliquishness

Moreover, this model shows how pure altruism i.e. unconditional cooperative behavior, is possible in social networks despite the penalties imposed by the social dilemma. Contrary to most models that explain cooperation in terms of cliquishness (often using a choice and refusal mechanism) we showed how altruist behaviors become sustainable thanks to the emergence of a social networks structure where authorities are altruist agents while hubs are reciprocators and selfish agents. Altruists can "survive" because they are more popular and have a more intense social life.

### 4.4 Perspectives about emergence of cooperation

To cite a few, further developments of this model should include errors on the action implementation since it is well known that this kind of errors is detrimental to cooperation. However, we conjecture that while errors will certainly decrease the rate of cooperation in a population, cooperation will still be sustainable for a large domain of the parameters space if the level of errors is no too high. Indeed, since agents can interact only once as a first mover, the only important thing for cooperation is that altruists and reciprocators have a non-empty address book. Because of errors in the partner selection process, agents are bound to meet several cooperators so that their address book quickly has few names. Errors on action implementation will certainly shorten this list of relations but if not too frequent, they should not qualitatively change our results. However, the interplay between error level on action, error level on partner selection and cooperation is highly none trivial and is worth being investigated.

# Mathematical Appendix

## I. Proof of proposition 1:

We will note $\Delta = \{(\gamma, \delta) | \delta + \gamma \leq 1 ; \delta \geq 0 \ \& \ \gamma \geq 0 \}$ and $n_i$ the number of agents of type $i$.

To compute to mean payoffs of each type we need first to compute the probability for an agent to be connected.

***Lemma 1*** : Let $i$ be an agent taken at random in a sub-population $P$ of agents which are looking for some particular type(s), in proportion $\lambda$ in the population, to do relations; the probability that $i$ has already done some relations is $P_{connect}(\lambda) = 1 - \dfrac{1-\theta}{1-(1-\lambda)\theta}$

**Proof :**

Let $i$ be an agent of age $t$ in $P$. The probability it hasn't met any agent to do relations during its life is $(1-\lambda)^t$. Consequently, the probability that agent $i$ of age $t$ has a non empty address book is $1-(1-\lambda)^t$. (We neglect here the occurrences when your only relation dies before you meet an other one, which will be a good approximation as soon as $e$ is not too small).

The distribution of ages in the population has the generatrice function $f(z) = (1-\theta)\sum_t \theta^t z^t$. Then the probability for an agent to be connected has the generatrice function $\phi(z) = \left(1 - \dfrac{1-\theta}{1-(1-\lambda)\theta}\right)z + 1 - \dfrac{1-\theta}{1-(1-\lambda)\theta}\lambda$. We deduce the mean connectedness with agents of the considered types(s): $P_{connect}(\lambda) = 1 - \dfrac{1-\theta}{1-(1-\lambda)\theta}$ and its variance $Var_{connect}(\lambda) = \left(1 - \dfrac{1-\theta}{1-(1-\lambda)\theta}\right)\dfrac{1-\theta}{1-(1-\lambda)\theta}$. ●

In the area $p/r<\delta$, the domain concerned by emergence of cooperation, only altruists cooperate first move. Consequently, they are the only one to be able to do relations with altruist *and* reciprocators. Their probability to be connected is then $P_{connect}(\delta+\gamma)$. Selfish agents and reciprocators, on the contrary, can only do relations with altruists. Their probability to be connected is then $P_{connect}(\gamma)$. We can now compute the payoffs.

Payoffs terms are composed of two independent terms: the expected payoffs as first mover and the expected payoffs as second mover. For each of them we have to treat separately the cases where the agent is connected, the cases where it is not and exploration moves. We then get:

$\pi_{selfish} = P_{connect}(\gamma)[(1-e)+e.p(1-\gamma)+e.\gamma] + ...$  # 1

$(1-P_{connect}(\gamma))[p.(1-\gamma)+\gamma] + ...$  # 2

$e.[P_{connect}(\gamma).(1-\gamma).p + P_{connect}(\gamma+\delta).\gamma] + ...$  # 3

$(1-P_{connect}(\gamma)).(1-\gamma).p + (1-P_{connect}(\gamma+\delta)).\gamma$  # 4

With :

# 1 : $i$ is connected, 1$^{rst}$ move interactions

# 2 : $i$ is not connected, 1$^{rst}$ move interactions at random

# 3 : 2$^{nd}$ move interactions with connected agents

# 4 : 2$^{nd}$ move interactions with not connected agents

After simplifications we get:

$$\pi_{selfish} = (1-e)\left(\frac{2p(1-\gamma+\theta(\gamma-1))+\gamma}{1+\theta(\gamma-1)} + \frac{(1-\theta)\gamma}{1-(1-(\delta+\gamma))\theta}\right) + 2e(p(1-\gamma)+\gamma) \ (\pi_s) \quad \text{(Eq. 1)}$$

In the same way :

$\pi_r = P_{connect}(\gamma)((1-e).r + e.(\gamma + p(1-\gamma))) + ...$  # 1

$(1-P_{connect}(\gamma))(p.(1-\gamma)+\gamma) + ...$  # 2

$e.(P_{connect}(\gamma).(1-\gamma).p + P_{connect}(\gamma+\delta).[(1-e).r.\gamma/(\gamma+\delta)+e.\gamma.r]) + ...$  # 3

$(1-P_{connect}(\gamma)).(1-\gamma).p + (1-P_{connect}(\gamma+\delta)).\gamma.r$  # 4

After simplification we get :

$$\pi_{reciprocators}=(1-e)\left(\frac{\theta\gamma(r+2p-1)+\gamma+2p(1-\gamma-\theta)}{1+\theta(\gamma-1)}+\frac{r\gamma}{1-(1-(\delta+\gamma))\theta}\right)+e(2p+\gamma(1-2p+r)) \quad (\pi_r) \quad \text{(Eq. 2)}$$

and

$\pi_a = P_{connect}(\gamma+\delta)[(1-e).r+e.r.(\gamma+\delta)]+...$  # 1

$(1- P_{connect}(\gamma+\delta)).(\delta+\gamma).r+...$  # 2

$P_{connect}(\gamma).(1-e).r.\delta/\gamma+P_{connect}(\gamma+\delta).[(1-e).r.\gamma/(\gamma+\delta)+e.\gamma.r]+...$  # 3

$(1-P_{connect}(\gamma+\delta)).\gamma.r$  # 4

After simplifications we get:

$$\pi_{altruists}=(1-e)\left(\frac{r\theta\delta}{1+\theta(\gamma-1)}+\frac{r(2\gamma+\delta)}{1-(1-(\delta+\gamma))\theta}\right)+er(\delta+2\gamma) \quad (\pi_a) \quad \text{(Eq. 3)}$$

## II. Theoretical activity

To compute the theoretical mean activity, we follow the same strategy as for the expected payoffs with the difference that we have to take in consideration interactions for which payoffs are *0*.

*Mean activity for selfish agents:*

Mean activity of agents can be split into two independents terms: the mean activity as first mover and the mean activity as second mover. For each of them we have to treat separately the cases where the agent is connected, the cases where it is not and exploration moves. We then get:

$\pi_{selfish}= 1$  # 1+ # 2

$e.[P_{connect}(\gamma).(1-\gamma)+ P_{connect}(\gamma+\delta).\gamma]+...$  # 3

$(1-P_{connect}(\gamma)).(1-\gamma)+(1- P_{connect}(\gamma+\delta)).\gamma$  # 4

With :

# 1 : *i* is connected, 1$^{rst}$ move interactions

# 2 : *i* is not connected, 1$^{rst}$ move interactions at random

*# 3 : 2$^{nd}$ move interactions with connected agents*

*# 4 : 2$^{nd}$ move interactions with not connected agents*

In the same way for reciprocators:

$\pi_r = 1 +$                                                                     *#1 + # 2*

   $e.P_{connect}(\gamma).(1-\gamma) + P_{connect}(\gamma+\delta).[(1-e).\gamma/(\gamma+\delta)+e.\gamma] + ...$        *# 3*

   $(1-P_{connect}(\gamma)).(1-\gamma)+(1-P_{connect}(\gamma+\delta)).\gamma$                 *# 4*

and for altruists :

In the same way for reciprocators :

$\pi_a = 1 +$                                                                     *#1 + # 2*

   $P_{connect}(\gamma).[e.(1-\gamma)+(1-e).\delta/\gamma] + P_{connect}(\gamma+\delta).[e.\gamma+(1-e).\gamma/(\gamma+\delta)] + ...$    *# 3*

   $(1-P_{connect}(\gamma)).(1-\gamma)+(1-P_{connect}(\gamma+\delta)).\gamma$                 *# 4*

## Complements

For the domain $\frac{p}{1-p} < \delta$, things are a little more complex. Since this area is not concerned by emergence of cooperation, we will only give some guidelines, with highlight on differences with the case treated above. As selfish agent cooperates first move, it may have a successful interaction with an altruist or a reciprocator. But the expected payoff associated with an altruist relation is higher than the one associated with a reciprocator. In fact, if a selfish agent *i* knows an altruist *j*, *i* will defect with *j* every times *i* will have the choice to do so. Therefore; the expected payoffs are $(1-e)\frac{1}{(1-\theta^2)}$ On the other hand, expected payoffs associated with a reciprocator relation are $(1-e)\frac{1-p}{(1-\theta^2)}$. It could be in its interest for a selfish agent to check if a given new relation is an altruist by defecting at the second interaction. Let us compute the expected payoffs of these two strategies:

1. S1: If it doesn't check, the expected payoffs are $1 - p + (1-e)\theta^2 \frac{1-p}{(1-\theta^2)}$

2. S2: If it does check, the expected payoffs are:

$$\frac{\delta}{\delta + \gamma}\left(p + (1-e)\theta^2 \frac{1-p}{(1-\theta^2)}\right) + \frac{\gamma}{\delta + \gamma}\left(1 + (1-e)\frac{\theta^2}{(1-\theta^2)}\right)$$

The difference is then :

$$S2 - S1 = \frac{\delta}{\delta + \gamma}(2p-1) + \frac{\gamma}{\delta + \gamma}\left(p + (1-e)\frac{\theta_2}{(1-\theta_2)}p\right)$$

$$S2 - S1 \geq 0 \Leftrightarrow \delta \frac{(2p-1)}{p\left(1 + (1-e)\frac{\theta_2}{(1-\theta_2)}\right)} \leq \gamma \qquad (Eq. 3)$$

The coefficient of $\delta$ in Eq. 3 is very low and consequently, $S2 - S1 \geq 0$ is true for almost the whole area $\frac{p}{1-p} < \delta$. We will expose here only the case $S2-S1>0$.

Then we have to compute for the special case of selfish agents, probability of finding the different types of address books: 1) empty, 2) with reciprocator(s) relations but no altruist relation, 3) with altruist(s) relations. After that, with the same type of calculus as in 2.1, we find that in simplified form :

$$\pi_{egoists} = (1-e)\left(\frac{2p(1-\gamma+\theta(\gamma-1))+\gamma}{1+\theta(\gamma-1)} + \frac{(1-\theta)\gamma}{1-(1-(\delta+\gamma))\theta}\right) + 2e(p(1-\gamma)+\gamma)$$

$$\pi_{reciprocators} = (1-e)\left(\frac{\theta p\gamma + 2p(1-\gamma-\theta)}{1+\theta(\gamma-1)} + \frac{(1-p)\gamma}{1-(1-(\delta+\gamma))\theta}\right) + e(2p+2\gamma-3p\gamma)$$

$$\pi_{altruists} = (1-e)\left(\frac{(1-p)\theta\delta}{1+\theta(\gamma-1)} + \frac{(1-p)(2\gamma+\delta)}{1-(1-(\delta+\gamma))\theta}\right) + e(\delta(1-p)+2\gamma-2p\gamma)$$

## *Computational Appendix*

We give here the algorithm used for the simulations.

**Parameters and notation**

The parameters are:

1. $e \in [0,1]$ : Exploration parameter

2. $p \in [0, \frac{1}{2}]$ : Game parameter

3. $\theta \in [0, \frac{1}{2}]$ : Parameter defining the mean lifetime expectancy.

The population at time *t* is noted *P(t)*.

**Initial conditions :**

$N$ agents with empty address books from the three types: altruist, reciprocator and selfish, in proportion $\delta$, $\gamma$ and $(1-\gamma-\delta)$.

**At each period of time t :**

*Set all payoffs to 0*

*Interaction (parallel updating):*

1. For each agent of *P(t)*, an individual binomial random variable defines if it will be allowed with a probability *1-e* to chose its partner. If its address book is empty, the agent has to interact with an unknown partner as first mover.

2. In case of interaction as first mover with an unknown partner, the partner is chosen at random and table 1 gives the action played.

3. In case of interaction as first mover in address book:
   - Altruists and reciprocators choose a relation at random and play *C*.
   - Selfish look for an altruist and defect. If they know only reciprocators, they choose one at random and cooperate. In they don't know altruists but know agents with ambiguous types (a (C,C) result in the preceding interaction as first mover), they choose one of them at random and defect to test whether it is an altruist. (This situation only happens in the area $\frac{p}{1-p} < \delta$ ).

4. Each agent replies as second mover according to its type (cf. table 1).

5. Each first mover adds the name of its second mover partner in its address book if the partner has cooperated. Selfish have special categories: a *(D,C)* results - *altruist,* a *(C,C)* result – *reciprocator or altruist*, a *(D,D)* result after interaction with *reciprocator or altruist* – *reciprocator.*

*Selection :*

1. After interactions took place, the total payoffs of each agent are computed by adding the payoffs of all two persons games they were engaged in this period.

2. Each agent dies with a probability $(1-\theta)$.

3. Died agents are replaced with new agents with empty address books. Died agents are suppressed from the address books of their friends. This lead to populations *P(t+1)*.

4. Types of new agents are determined by a replicator dynamics indexed on the total payoffs of agents from *P(t)*: for each new born agent, an agent of *P(t)* is selected with a probability proportional to its total payoffs. The new agent inherits the type of this particular agent.